\documentclass{article}
\usepackage{arxiv}
\usepackage[utf8]{inputenc} 
\usepackage[T1]{fontenc}    
\usepackage{hyperref}       
\usepackage{url}            
\usepackage{booktabs}       
\usepackage{amsfonts}       
\usepackage{nicefrac}       
\usepackage{microtype}      
\usepackage{lipsum}		
\usepackage{graphicx}
\usepackage{natbib}
\usepackage{doi}
\usepackage{bm}
\usepackage{mathtools}
\usepackage{amsmath}
\usepackage{braket}
\usepackage{xcolor}
\usepackage{amsthm}

\theoremstyle{definition}

\newtheorem{theorem}{Theorem}
\newtheorem{lemma}[theorem]{Lemma}

\DeclarePairedDelimiter\abs{\lvert}{\rvert}
\DeclarePairedDelimiter\norm{\lVert}{\rVert}
\DeclarePairedDelimiter\ceil{\lceil}{\rceil}
\DeclarePairedDelimiter\floor{\lfloor}{\rfloor}

\title{Detailed Error Analysis of the HHL Algorithm}

\author{ \href{https://orcid.org/0000-0000-0000-0000}{\includegraphics[scale=0.06]{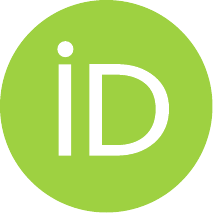}\hspace{1mm}Xinbo Li} \\
	Electrical and Computer Engineering\\
	University of Manitoba\\
	Winnipeg, MB, Canada  \\
	\texttt{lix34545@myumanitoba.ca} \\
	\And
	\href{https://orcid.org/0000-0000-0000-0000}{\includegraphics[scale=0.06]{orcid.pdf}\hspace{1mm}Christopher Phillips} \\
	Independent\\
	Brandon, MB, Canada \\
	\texttt{christopherdphillips7@gmail.com} \\
}

\renewcommand{\shorttitle}{\textit{arXiv} Template}

\hypersetup{
pdftitle={A template for the arxiv style},
pdfsubject={q-bio.NC, q-bio.QM},
pdfauthor={David S.~Hippocampus, Elias D.~Striatum},
pdfkeywords={First keyword, Second keyword, More},
}

\begin{document}
\maketitle

\begin{abstract}
We reiterate the contribution made by Harrow, Hassidim, and Llyod to the quantum matrix equation solver with the emphasis on the algorithm description and the error analysis derivation details.
Moreover, the behavior of the amplitudes of the phase register on the completion of the Quantum Phase Estimation is studied.
This study is beneficial for the comprehension of the choice of the phase register size and its interrelation with the Hamiltonian simulation duration in the algorithm setup phase.
\end{abstract}


\section{Introduction}
The seminal paper \cite{Harrow_2009} proposes the quantum matrix equation solver, commonly referred to as the HHL algorithm. 
Compared to classical matrix equation solvers, the HHL algorithm offers exponential speedup thanks to the expressive power of quantum information storage and processing.
For a matrix equation $A \mathbf{x} = \mathbf{b}$, the HHL algorithm seeks to prepare a quantum state $\ket{x}$ that is proportional to the desired solution vector $\bm{x}$. 
When the matrix $A$ is $s$-sparse, i.e., containing at most $s$ nonzero elements per row/column, the HHL algorithm completes in $\tilde{O}(\log (N)s^2 \kappa^2 /\epsilon)$ with a target error level $\epsilon$, where $\kappa$ is the condition number of $A$.
A conjugate gradient method \cite{shewchuk1994introduction}, would instead complete in ${O}(N s \kappa \log (1/\epsilon))$ in general (when $A$ is not necessarily positive definite).
Hence, the HHL algorithm achieves exponential speedup over $N$, even though containing a worse dependence on $s$, $\kappa$, and $\epsilon$.
The quadratic dependence on $s$ is acceptable assuming $s$ being a small number, and in a dense matrix equation case, the HHL is extended in \cite{Wossnig2018}. 
The worse dependence on the condition number and error is addressed by \cite{clader2013preconditioned} and \cite{Childs_2017}

Note that the outcome of the HHL algorithm is a quantum state, so subsequent quantum post-processing is assumed, otherwise the exponential speedup is voided if the entire vector needs to be transformed into a classical memory (writing down the solution would take $O(N)$ time).
One circumstance and important application in which the HHL speedup is preserved is when the expectation with respect to some observable $M$, i.e., $\mathbf{x}^\dagger M \mathbf{x}$ is of interest. 
In this case, such implementation of the observable as a quantum circuit at the end of the HHL circuit, and the expectation is obtained by measuring in an appropriate basis, a standard treatment in quantum algorithms \cite{nielsen2010quantum}.

As the matrix equation is a basic linear algebraic procedure, it is no surprise that the HHL algorithm underpins a collection of other quantum algorithms \cite{grant2018hierarchical, butler2018machine, biamonte2017quantum}, which again underscores the pivotal role of understanding the HHL algorithm for a quantum software engineer.
We note, however, that the HHL algorithm is not a near-term algorithm suitable for Noisy Intermediate-Scale Quantum (NISQ) devices \cite{preskill2018quantum}. 
Though research on near-term version of it has emerged \cite{yalovetzky2021nisq}, the study of the HHL algorithm mainly remains theoretical level.

Despite being well-presented in the original paper overall, in this work, the theoretical framework of the HHL algorithm is revisited with added elaborations to derivation and proof, and corrections on typos/mistakes.
The target of this work is to help readers understand the HHL algorithm in detail so that possible improvement can be instilled on this groundwork.
We confine our scope to the study of the original HHL algorithm regardless of previously mentioned improvements \cite{Wossnig2018, clader2013preconditioned, Childs_2017}.

The paper is structured as follows.
Section \ref{sec: algorithm detail} covers the detailed breakdown of the HHL algorithm with recorded intermediate results of each step.
Next, the behavior of the amplitudes of the phase register is studied in Section \ref{sec: amplitude behavior}.
This behavior is paramount in understanding the choice of the register size, as well as the study of the error.
In Section \ref{sec: error analysis}, a detailed error analysis of the HHL algorithm is given with proofs and derivations.

\section{Algorithm Details} \label{sec: algorithm detail}
Herein, we make the following assumptions for the matrix equation for simplicity: 
\begin{itemize}
    \item $A \in \mathbb{C}^{N \times N}$ is Hermitian, where $N$ is a power of 2 such that $n \coloneqq \log_2(N)$ is an integer. 
    \item The condition number $\kappa$ of $A$ is known or can be efficiently approximated, and $A$ is scaled so that all of its eigenvalues belong to the region $[1/\kappa, 1]$.
    \item $\bm{b}$ is a normalized vector so that it can be encoded into a quantum state $\ket{b}$. 
\end{itemize}


\begin{figure}[h]
\centering
\includegraphics[width=0.8\columnwidth]{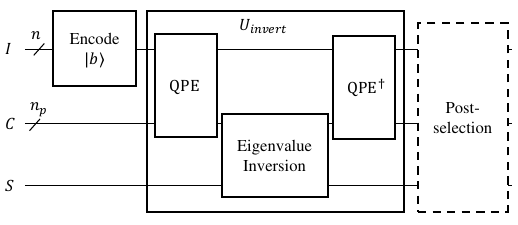}
\caption{The diagram illustration of the HHL circuit.}
\label{fig: 1. HHL circuit}
\end{figure}

The diagram of the HHL algorithm is given in Figure \ref{fig: 1. HHL circuit}.
The HHL circuit contains three registers: the Input/Output register $I$ with $n$ qubits; the Clock\footnote{Some recent sources name this register the "Phase Register".} register $C$ with $n_t$ qubits; the Flag register $S$ that contains a qutrit (a quantum entity that lives in a 3-dimensional space). 
We define $T \coloneqq 2^{n_t}$ as the dimension of the Hilbert space encapsulating the statevectors in register $C$.


We assume that some other procedure \cite{Zhang_et_al_state_preparation} encodes state $\ket{b}$ into the register $I$, such that the initial state that $U_{invert}$ operates on is  
\begin{align}
    \ket{\Phi_0} = \ket{b}_I \ket{0}_C \ket{\text{nothing}}_S
     = \sum_{j=0}^{N-1} \beta_j \ket{u_j}_I \ket{0}_C \ket{\text{nothing}}_S,
\end{align}
where $\ket{b}$ is represented as $\sum_{j=0}^{N-1} \beta_j \ket{u_j}$ in the eigenbasis $\{\ket{u_j}\}$ of $A$: $A\ket{u_j} = \lambda_j \ket{u_j}$.
The unitary operation $U_{invert}$ represents the HHL procedure that inverts the matrix $A$. 
It contains three components: quantum phase estimation (QPE), eigenvalue inversion, and the inverse of QPE.

The steps of a practical $U_{invert}$ are
\begin{enumerate}
\item QPE, denoted as a unitary operator $P$.
    \begin{enumerate}
        \item  Prepare the state $\ket{\Phi}=\sqrt{\frac{2}{T}}\sum_{\tau=0}^{T-1} \sin \left(\frac{\pi}{T} (\tau + \frac{1}{2}) \right) \ket{\tau} $ on $C$.
        The state after this step is 
        \begin{align}
            \ket{\Phi_1} = \sum_{j=0}^{N-1} \beta_j \ket{u_j}_I 
            \sqrt{\frac{2}{T}}\sum_{\tau=0}^{T-1} \sin \left[\frac{\pi}{T} \left(\tau + \frac{1}{2} \right) \right] \ket{\tau}_C
            \ket{\text{nothing}}_S.
        \end{align}
    
        \item Repeatedly apply the Hamiltonian simulation (HS) $e^{i A t_0/T}$ $\tau$ times to register $I$ conditioned on the state of the clock register $\ket{\tau}_C$, \footnote{Herein we use the subscript to denote the register for the states and operators, e.g., $U_I$ means a unitary operator $U$ applied to register $I$, and $\ket{u}_I$ means that register $I$ is in state $\ket{u}$.} an operation represented as $\sum_{\tau=0}^{T-1} e^{iA (t_0/T) \tau}_I \otimes \ket{\tau}\bra{\tau}_C$. 
        The choice of the HS time $t_0$ will be discussed in Section \ref{sec: amplitude behavior}. 
        The state of the circuit after the condition HS is
        \begin{align}
            \ket{\Phi_2} = \sum_{j=0}^{N-1} \beta_j \ket{u_j}_I 
            \sqrt{\frac{2}{T}}\sum_{\tau=0}^{T-1} \sin \left[\frac{\pi}{T} \left(\tau + \frac{1}{2} \right) \right] e^{i \lambda_j (t_0/T) \tau} \ket{\tau}_C
            \ket{\text{nothing}}_S.
        \end{align}

        \item Apply the quantum Fourier transform (QFT)
        \begin{align}
            \ket{\tau}_C \xrightarrow{\text{QFT}} \frac{1}{\sqrt{T}} \sum_{k=0}^{T-1} e^{-i (2\pi/T) k \tau} \ket{k}_C
        \end{align}
        to register $C$.
        The resultant state is 
        \begin{align} \label{eq: state after QPE}
            \ket{\Phi_3} = 
            \sum_{j=0}^{N-1} \beta_j \ket{u_j}_I
            \sum_{k=0}^{T-1} \alpha_{k|j} \ket{k}_C
            \ket{\text{nothing}}_S
        \end{align}
        where the amplitude is given as
        \begin{align} \label{eq: alpha}
            \alpha_{k|j} =
            \frac{\sqrt{2}}{T}\sum_{\tau=0}^{T-1} \sin \left[\frac{\pi}{T} \left(\tau + \frac{1}{2} \right) \right] e^{i \left(\lambda_j - \frac{2\pi}{t_0} k \right) \frac{t_0}{T} \tau} .
        \end{align}
        The phase in \eqref{eq: alpha} suggests the definition of the approximated eigenvalue (associated with state $\ket{k}_C$) as 
        \begin{align} \label{eq: approximated eigenvalue}
            \tilde{\lambda}_k \coloneqq \frac{2\pi}{t_0} k.
        \end{align}
        We further define $\delta_\lambda$ as the error in the eigenvalue approximation:
        \begin{align} \label{eq: delta lambda}
            \delta_\lambda (k|j) \coloneqq \lambda_j - \tilde{\lambda}_k.
        \end{align}
    \end{enumerate}
    At this point we have covered all three steps of the QPE.
    The idea of QPE is that the magnitude of amplitude, $|\alpha_{k|j}|$, peaks when $\abs{\delta_\lambda}$ is small, and decays when $\abs{\delta_\lambda}$ is large, such that the terms with $k$ that yields a good eigenvalue approximation dominates.
    This behavior is essential to understanding the algorithm and the error analysis of the algorithm.
    We will elaborate on this in Section \ref{sec: amplitude behavior}. 
    Discreet reader might also notice that the lower and upper bounds of $k$ are not given in \eqref{eq: state after QPE}, we will give these in Section \ref{sec: amplitude behavior} as well.
        
    \item Set the a flag register $S$ in the state 
    \begin{align} \label{eq: h(lambda) definition}
        &\ket{h(\tilde{\lambda}_k)} \coloneqq \nonumber \\
        &\quad \begin{cases}
        f(\tilde{\lambda}_k) \ket{\text{well}} + g(\tilde{\lambda}_k) \ket{\text{ill}} + \sqrt{1-f^2(\tilde{\lambda}_k) - g^2(\tilde{\lambda}_k)}\ket{\text{nothing}}, & k = 0, 1, \cdots, K \\
        \ket{\text{nothing}}, & k = K+1, \cdots, T-1
        \end{cases}
    \end{align} 
    where the filter functions $f(\lambda)$ and $g(\lambda)$ are defined as
    \begin{align}
        f(\lambda) \coloneqq 
        \begin{cases}
            \frac{1}{2 \tilde{\kappa} \lambda}, & \lambda \in \left[\frac{1}{\tilde{\kappa}} , 1 \right] \\
            -\frac{1}{2} \cos(\pi \tilde{\kappa} \lambda), & \lambda \in \left[ \frac{1}{2\tilde{\kappa}} , \frac{1}{\tilde{\kappa}} \right] \\
            0, & \lambda \in \left( 0,  \frac{1}{2\tilde{\kappa}} \right] 
        \end{cases} 
    \end{align}
    \begin{align}
        g(\lambda) \coloneqq
        \begin{cases}
            0, & \lambda \in \left[\frac{1}{\tilde{\kappa}} , 1 \right] \\
            \frac{1}{2} \sin(\pi \tilde{\kappa} \lambda), & \lambda \in \left[ \frac{1}{2\tilde{\kappa}}, \frac{1}{\tilde{\kappa}} \right] \\
            \frac{1}{2}, & \lambda \in \left( 0,  \frac{1}{2\tilde{\kappa}} \right] 
        \end{cases} 
    \end{align}
   where $\tilde{\kappa}$ is the approximation of the actual condition number $\kappa$ with the assumption that 
   \begin{align} \label{eq: kappa_t and kappa}
        \tilde{\kappa} = O(\kappa).     
   \end{align}
   The cutoff value $K$ in \eqref{eq: h(lambda) definition} relates to the choice of $t_0$ and $T$, and will be discussed in the next section. 
    The state after this step is 
    \begin{align} \label{eq: state before QPE inverse, approximated}
        \ket{\Phi_4} = 
        \sum_{j=0}^{N-1} \beta_j \ket{u_j}_I
        \sum_{k=0}^{T-1} \alpha_{k|j} \ket{k}_C
        \ket{h(\tilde{\lambda}_k)}_S.
    \end{align}

    \item Apply inverse QPE $P^\dagger$, i.e., apply the inverse of the three steps of QPE in reverse order to restore the $\ket{0}$ state in register $C$.
    The final state $\ket{\Phi_f}$ would be 
    \begin{align} \label{eq: final state, approximated}
        \ket{\Phi_f} = P^\dagger \sum_{j=0}^{N-1} \beta_j \ket{u_j}_I
        \sum_{k=0}^{T-1} \alpha_{k|j} \ket{k}_C
        \ket{h(\tilde{\lambda}_k)}_S.
    \end{align}
\end{enumerate}
The final state \eqref{eq: final state, approximated} from the practical HHL circuit $QC_{practical}$ does not directly relate to the solution of the matrix equation $\bm{x} = \sum_{j = 0}^{N-1} \frac{\beta_j}{\lambda_j} \ket{u_j}$ in an obvious way.
This is because the non-ideal QPE, which does not guarantee that $|\alpha_{k|j}| = 0$ when $|\delta_\lambda (k|j)| \neq 0$, results in the entanglement between the clock register $C$ and flag register $S$ in \eqref{eq: state before QPE inverse, approximated}. 
This entanglement will still take place in the final state $\ket{\Phi_f}$, making its comparison with $\ket{x}$ hard.

To offer intuition on why HHL algorithm works, we consider an ideal HHL circuit $QC_{ideal}$, in which we suppose that the QPE reveals the exact eigenvalue.
Let us use a overhead bar to distinguish the circuit components and quantum states of $QC_{ideal}$ from those of $QC_{practical}$.
For example, $\Bar{P}$ means the ideal QPE in $QC_{ideal}$.
The final state of $QC_{ideal}$, $\ket{\bar{\Phi}_f}$ would be
\begin{align} \label{eq: final state, ideal}
    \ket{\bar{\Phi}_f} = \sum_{j=0}^{N-1} \beta_j \ket{u_j}_I \ket{0}_C \ket{h(\lambda_j)}_S.
\end{align}

The comparison between \eqref{eq: final state, ideal} and \eqref{eq: h(lambda) definition} indicates that if we post-select the final state depending on the state of the flag register $S$, we will retrieve a solution state $\ket{\bar{x}}$ that is proportional to the solution to the matrix equation $\bm{x}$, provided that all eigenvalues are in the well-conditioned region $[\frac{1}{\tilde{\kappa}}, 1]$. 

In Section \ref{sec: error analysis}, we will prove that $\norm {\ket{\Phi_f} - \ket{\bar{\Phi}_f}} = O(\kappa/t_0)$. 
Moreover, the distance between the non-ideal solution state $\ket{x}$ post-selected from $\ket{\Phi_f}$ and the ideal solution statte $\ket{\bar{x}}$ is also upper bounded by $O(\kappa/t_0)$. 
Hence, $QC_{practical}$ can produce a desired solution up to some error level with properly chosen $t_0$.

\section{The Behavior of the Amplitudes $\abs{\alpha_{k|j}}$} \label{sec: amplitude behavior}

We define $\delta \coloneqq t_0 \delta_\lambda = \lambda_j t_0 - 2\pi k$, then the magnitude of the amplitude can be simplified as 
\begin{align} \label{eq: amplitude magnitude}
    \abs{\alpha_{k|j}} = \frac{\sqrt{2}}{T}   \sin{\left(\frac{\pi}{2T}\right)}
    \frac{ \abs*{\cos{(\frac{\delta}{2T})} \cos(\frac{\delta}{2})}}{\abs*{\sin{\left(\frac{\delta+\pi}{2T} \right)}\sin{\left(\frac{\delta - \pi}{2T} \right)}}} .
\end{align}
The target of QPE, is to make $\abs{\alpha_{k|j}} \to 1$ when $\abs{\delta_\lambda} \to 0$ (corresponding to a good eigenvalue approximation), and $\abs{\alpha_{k|j}} \to 0$ when $\abs{\delta_\lambda}$ is large (corresponding to a poor eigenvalue approximation).
In \cite{Harrow_2009}, this is descried by a desired upper bound $\frac{8\pi}{\delta^2}$ for the amplitude $\abs{\alpha}$.
The attainability of this target is contingent upon two critical hyperparameters in QPE: $t_0$ and $T$, as indicated from \eqref{eq: approximated eigenvalue} and \eqref{eq: amplitude magnitude} that they are critical in deciding the approximated eigenvalue and the corresponding eigenvalue approximation quality.
In this section, we establish the selection criteria for $t_0$ and $T$ by examining the behavior of $\abs{\alpha_{k|j}}$ in relation to these choices.

Suppose we use the entire clock register state range, i.e., $k = 0, 1, 2, \cdots, T-1$, then the range of estimable eigenvalues is 
\begin{align}
R\{\tilde{\lambda}_k\} = \left[ \frac{2\pi}{t_0},
\frac{2\pi(T-1)}{t_0}  \right].
\end{align}
For mathematical formulation simplicity, hereafter within this section we modify the assumption on the true range of the eigenvalues from $[1/\kappa, 1]$ 
\begin{align}
R\{\lambda_j\} = \frac{T-1}{T}\left[\frac{1}{\kappa}, 1 \right] .  
\end{align}
Notice that this change would only require a re-scaling of the original matrix in practice.

One obvious requirement on $t_0$ and $T$ for a successful eigenvalue approximation is that $R\{\lambda_j\} \subset R\{\tilde{\lambda}_k\}$, in other words, conditions 
\begin{align} \label{eq: condition largest eigval}
    \frac{2\pi(T-1)}{t_0} \geq \frac{T-1}{T} 
    \qquad \Rightarrow \qquad
    t_0 \leq 2\pi T
\end{align}
and 
\begin{align} \label{eq: condition smallest eigval}
    \frac{2\pi}{t_0} \leq \frac{T-1}{T} \frac{1}{\kappa}
\end{align}
need to be satisfied.

We first study the simpler condition \eqref{eq: condition largest eigval}. 
Let us choose the boundary case $t_0 = 2\pi T$, with which the maximum estimable eigenvalue happens to be the largest true eigenvalue $\frac{T-1}{T}$.
With such choice, \eqref{eq: condition smallest eigval} yields $T \geq \kappa +1$.
However, note that $\abs{\alpha}$ is a continuous $2\pi T$-periodic function with respect to $\delta$, which indicates that $k=0$ and $k=T-1$ gives similar amplitude magnitudes.
Namely, $\abs{\alpha(\delta(k=0))} \approx \abs{\alpha(\delta(k=T-1))}$, because the input difference $\delta(k=T-1) - \delta(k=0) = 2\pi (T-1)$ is close to one period.
This periodic behavior of $\abs{\alpha}$ is not detrimental when the target eigenvalue $\lambda_j$ is around the center of $R\{\lambda_j\}$, but when approximating small and larger eigenvalues (eigenvalues that are close to $\frac{T-1}{T} \frac{1}{\kappa}$ and $\frac{T-1}{T}$), the periodicity of $\abs{\alpha}$ causes unwanted large amplitude at the corresponding other end of the approximated spectrum. 
Figure \ref{fig: 2. |alpha| versus delta for various eigenvalues, t0=2piT} illustrates this with three instances of eigenvalues $\frac{T-1}{T}\frac{1}{\kappa}, \frac{T-1}{2T}$, and $\frac{T-1}{T}$, exemplifying small, moderate, and large values respectively.
In this figure, we choose $t_0 = 2\pi T, T = \kappa+1$.
The desired upper bound for the amplitude \cite{Harrow_2009} $\frac{8\pi}{\delta^2}$ is shown to demonstrate the region where this upper bound is violated.
The continuous periodic pattern of $\abs{\alpha}$ can be perceived from Figure \ref{fig: 2. |alpha| versus delta for various eigenvalues, t0=2piT}: the black curve in each sub-figure manifests at the opposite extremity. 
This is because QPE captures the periodic nature of phase: $0$ and $2\pi$ are in fact the same point in the polar plane.
When the actual eigenvalue is close to the left boundary $\frac{T-1}{T}\frac{1}{\kappa}$ or the right boundary $\frac{T-1}{T}$, large amplitudes appear at the poor eigenvalue approximation range, i.e., when $\abs{\delta}$ is large. 
As a consequence, the desired upper bound is violated in the unwanted region of the approximated eigenvalue spectrum.
\begin{figure}[h]
    \centering
    \includegraphics[width=0.8\columnwidth]{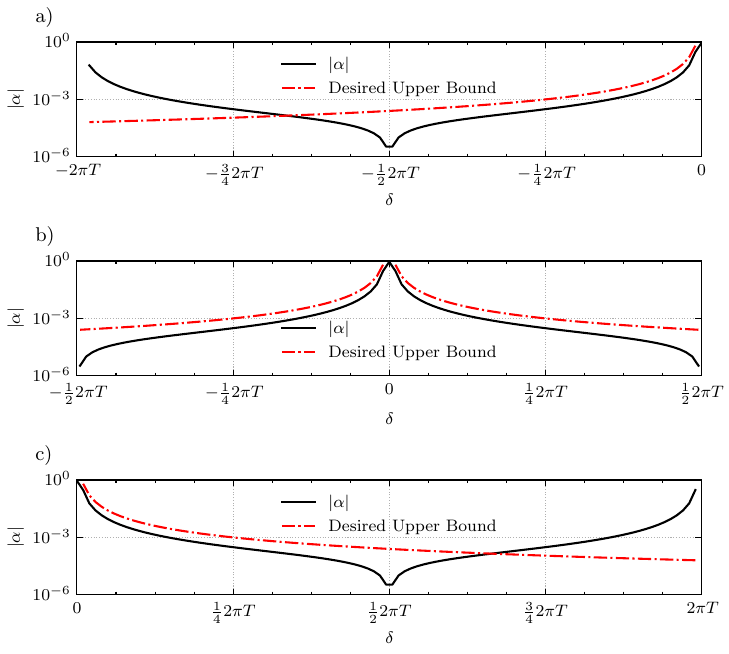}
    \caption{$\abs{\alpha}$ versus $\delta$ for a) small eigenvalue $\frac{T-1}{T} \frac{1}{\kappa}$, b) moderate eigenvalue $\frac{T-1}{2T}$, and c) large eigenvalue $\frac{T-1}{T}$ with $t_0 = 2\pi T, T = \kappa + 1.$}
    \label{fig: 2. |alpha| versus delta for various eigenvalues, t0=2piT}
\end{figure}

This issue can be avoided by choosing a $t_0$ strictly smaller than $2\pi T$, preserving \eqref{eq: condition largest eigval}, a maneuver with the side effect of wasting some range of the clock register.
To explain, let us consider the choice of $t_0 = \pi T$, in which case the approximated eigenvalues are 
\begin{align}
    \tilde{\lambda}_k = 2\frac{k}{T} = 0, 2\frac{1}{T}, 2\frac{2}{T}, \cdots.
\end{align}
Hence, the values of $k$ from $0$ to $\ceil*{(T-1)/2}$ is sufficient to cover the entire range of the true eigenvalues.
For the problematic smallest and largest eigenvalues which made $\abs{\alpha}$ violate the desired upper bound in Figure \ref{fig: 2. |alpha| versus delta for various eigenvalues, t0=2piT}, with the new choice of $t_0 = \pi T$, when $k$ is in the range from $\ceil*{(T-1)/2}$ to $T-1$, the plot of $\abs{\alpha}$ folds to the symmetric counterpart, confined to the desired upper bound. 
This behavior of $\abs{\alpha}$ versus $\delta$ is visualized in Figure \ref{fig: 3. |alpha| versus delta for various eigenvalues, t0=piT} with the same three examples of $\lambda_j$ as in Figure \ref{fig: 2. |alpha| versus delta for various eigenvalues, t0=2piT}.
Note that in addition to $t_0 = \pi T$ satisfying condition \eqref{eq: condition largest eigval}, condition \eqref{eq: condition smallest eigval} requires that $T \geq 2\kappa +1$.
In Figure \ref{fig: 3. |alpha| versus delta for various eigenvalues, t0=piT}, this choice is made as $T = \ceil*{2\kappa +1}$.
One can observe that the desired upper bound is not violated in the entire range of $k$, regardless of the actual eigenvalue.

\begin{figure}[h]
    \centering
    \includegraphics[width=0.8\columnwidth]{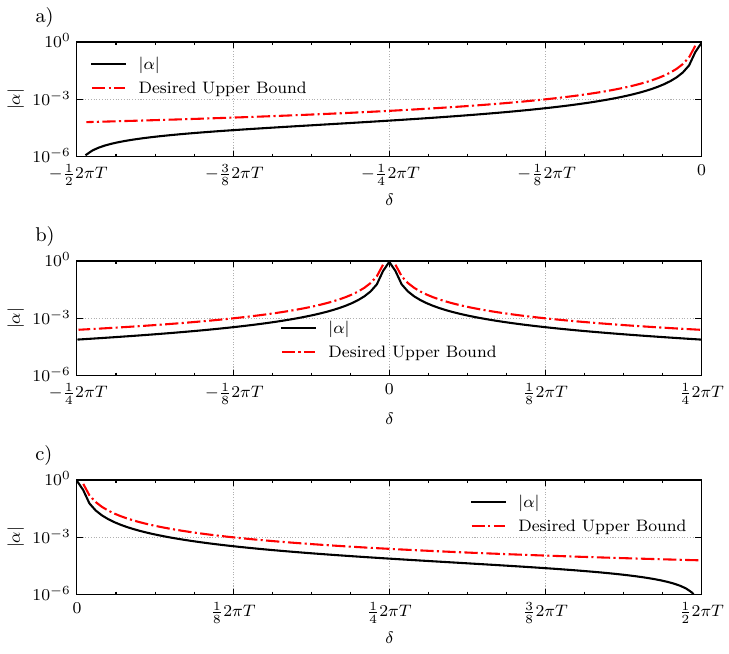}
    \caption{Data similarly illustrated as in Figure \ref{fig: 2. |alpha| versus delta for various eigenvalues, t0=2piT} with $t_0 = \pi T, T = 2\kappa +1$.}
    \label{fig: 3. |alpha| versus delta for various eigenvalues, t0=piT}
\end{figure}

Choosing $t_0 = \pi T$ can be effectively regarded as that the clock register range is not fully used.
This ``waste'' of the clock register is necessary to truncate half of the period of $\abs{\alpha}$, removing the unwanted tail on the opposite side when the actual eigenvalue $\lambda_j$ is close to either side of the boundary.
More generally, choosing $t_0 = \gamma (2\pi T)$ with $\gamma < \frac{1}{2}$ will utilize less spectrum while truncating the $\abs{\alpha}$ more. 
A smaller $\gamma$ results in a smaller $t_0$, namely a faster Hamiltonian simulation \cite{berry2007efficient, childs2010relationship}, hence would be favorable. 
However, note that a small $\gamma$ would result in a larger $T$: according to \eqref{eq: condition smallest eigval}, $T \geq \kappa/\gamma + 1$.
In other words, a shorter Hamiltonian simulation comes at a cost of requiring a larger clock register. 

In summary, the above analysis on the amplitude behavior suggests that $t_0$ is $T$ multiplying a coefficient, where $T$ has lower bound inversely proportional to this coefficient, and proportional to the condition number. 
In the next section of the error analysis, which is built on the amplitude analysis in this section.
The error will be proved to be bounded by $O(\kappa/t_0)$.
Hence, in practice, to achieve an error level of $\epsilon$, $t_0$ is chosen according to $O(\kappa/\epsilon)$ and so is $T$.
When the matrix is ill-conditioned, a large clock register size $T$ and long Hamiltonian simulation time $t_0$ is required to maintain accuracy.

\section{Error Analysis} \label{sec: error analysis}
The supplemental material accompanied with the original paper \cite{Harrow_2009} established a through error analysis of the error propagation of the algorithm (though with some critical typos and misuses of notations that may cause confusion). 
We perform a rigrous error analysis in this section following the framework presented in \cite{Harrow_2009}.

Since we have defined $QC_{ideal}$ and $QC_{practical}$ in Section \ref{sec: algorithm detail}, we can use $QC_{ideal}$ as a benchmark to evaluate the error behavior of $QC_{practical}$.
As a preview, the claims we make from the error analysis are summarized as the following theorem:
\begin{theorem}[The HHL Error Bound] \label{theorem: error bound}
\begin{enumerate}
    \item The error between the ideal final state $\ket{\bar{\Phi}_f}$ and the actual final state $\ket{\Phi}_f$ is bounded as \footnote{The first claim in \cite{Harrow_2009} is that the operator }
    \begin{align} \label{eq: error bound, no post-selection}
        \norm{\ket{\Phi_f}  - \ket{\bar{\Phi}_f}} = O(\kappa/t_0).
    \end{align}

    \item If we post-select on the flag register being in the space spanned by $\{\ket{well}, \ket{ill}\}$, and define the normalized state of register $I$ as $\ket{\bar{x}}$ in $QC_{ideal}$ and $\ket{x}$ in $QC_{practical}$, then 
    \begin{align} \label{eq: error bound, post selection to well and ill}
        \norm{\ket{x}  - \ket{\bar{x}}} = O(\kappa/t_0).
    \end{align}
    
    \item If $\ket{b}$ belongs to the well-conditioned subspace of $A$, and we post-select on $\ket{well}$ state of the flag register, then with the same definitions for $\ket{\bar{x}}$ and $\ket{x}$ as in the last case,
    \begin{align} \label{eq: error bound, post selection to well}
        \norm{\ket{x}  - \ket{\bar{x}}} = O(\kappa/t_0).
    \end{align}
    
\end{enumerate}
\end{theorem}

We use the next three subsections to prove the three claims in Theorem \ref{theorem: error bound}.

\subsection{Proof of \eqref{eq: error bound, no post-selection}: Error Bound in the Final State Without Post-Selection} \label{sec: error bound, no post-selection}

From linear algebra, the $L_2$ norm of the difference between any two arbitrary quantum states $\ket{\varphi_1}$ and $\ket{\varphi_2}$ is computed as
\begin{align} \label{eq: distance between two states}
    \norm{ \ket{\varphi_1} -\ket{\varphi_2} } = \sqrt{2(1-\Re\{\braket{\varphi_1|\varphi_2}\})}. 
\end{align}
In particular, 
\begin{align} \label{eq: distance between two ideal and practical final state, no post selection}
    \norm{ \ket{\Phi_f} -\ket{\bar{\Phi}_f} } = \sqrt{2(1-\Re\{\braket{\Phi_f|\bar{\Phi}_f}\})}
\end{align}
Hence, finding the upper bound of the error $ \norm{\ket{\Phi_f}  - \ket{\bar{\Phi}}_f}$ amounts to finding the lower bound of $\Re \{\braket{\Phi_f|\bar{\Phi}_f}\}$.
We analyze the inner product as follows.
 
Note that \eqref{eq: final state, approximated} is inverse QPE operating on \eqref{eq: state before QPE inverse, approximated}:
\begin{align} \label{eq: final state, QPE inverse on Phi4, approximated}
    \ket{\Phi_f} = P^\dagger \sum_{j=1}^N \beta_j \ket{u_j}_I \sum_{k=0}^{T-1} \alpha_{k|j} \ket{k}_C \ket{h(\tilde{\lambda}_k)}_S. 
\end{align}

From \eqref{eq: final state, QPE inverse on Phi4, approximated} and \eqref{eq: final state, ideal} and some manipulations, one can derive \footnote{See Appendix \ref{appd: inner product} for details} that 
\begin{align} \label{eq: inner product, final states}
    \braket{\Phi_f|\bar{\Phi}_f} 
    = 
    \sum_{j=0}^{N-1} \sum_{k=0}^{T-1} \abs{\beta_j}^2 \abs{\alpha_{k|j}}^2 \braket{h(\tilde{\lambda}_k)|h(\lambda_j)} \in \mathbb{R}.
\end{align}
To explain why this inner product is a real number, we note that as the matrix $A$ is Hermitian, its eigenvalue $\lambda_j \in \mathbb{R}$. 
The approximated eigenvalue $\tilde{\lambda}_k \in \mathbb{R}$ according to \eqref{eq: approximated eigenvalue}. 
The mapping $f, g$ are both $\mathbb{R} \to \mathbb{R}$, which renders every amplitude in $\ket{h(\lambda)}$ real. 
As a result, $\braket{h(\tilde{\lambda}_k)|h(\lambda_j)}\in \mathbb{R}$, and \eqref{eq: inner product, final states} is real.
The real part of the inner product \eqref{eq: inner product, final states} can be separated into two parts:
\begin{align} \label{eq: Re(inner product)}
    \Re\{\braket{\Phi_f|\bar{\Phi}_f} \}
    =
    \braket{\Phi_f|\bar{\Phi}_f} 
    = \underbrace{\sum_{j=0}^{N-1} \sum_{k: \abs{\delta} \leq 2\pi}  \abs{\beta_j}^2 \abs{\alpha_{k|j}}^2 \braket{h(\tilde{\lambda}_k)|h(\lambda_j)} }_\text{Term 1} + 
    \underbrace{\sum_{j=0}^{N-1} \sum_{k: \abs{\delta} > 2\pi} \abs{\beta_j}^2 \abs{\alpha_{k|j}}^2 \braket{h(\tilde{\lambda}_k)|h(\lambda_j)}}_\text{Term 2}.
\end{align}
according to the condition 
\begin{align} \label{eq: condition}
    k: \abs{\delta} \leq 2 \pi \Longleftrightarrow  
    \abs{k - \frac{t_0}{2\pi} \lambda_j} \leq 1.
\end{align}
Condition \eqref{eq: condition} being satisfied means that the approximated eigenvalue $\tilde{\lambda}_k$ is in the direct proximity of the true eigenvalue $\lambda_j$: only values of $k$ that satisfy \eqref{eq: condition} are $\ceil*{\frac{t_0}{2\pi} \lambda_j}$, $\floor*{\frac{t_0}{2\pi} \lambda_j}$, and possibly $\frac{t_0}{2\pi} \lambda_j$ itself if it is an integer. 
Hence, Term 1 and Term 2 correspond to the error contribution from good and poor eigenvalue approximations, respectively.

Next, we seek the lower bounds for both Term 1 and Term 2 to lower bound $\Re{\braket{\tilde{\varphi}|\varphi}}$.
To this end, we need to find the lower bound for $\Re\{\braket{h(\tilde{\lambda}_k)|h(\lambda_j)}\} = \braket{h(\tilde{\lambda}_k)|h(\lambda_j)}$, a result that can be directly derived from the following Lemma \ref{lemma: continuity}.

\begin{lemma}[The continuity of the mapping $\lambda \to \ket{h(\lambda)}$\footnote{Short proof of this Lemma: Note that $h(\lambda)$ is continuous in $[1/\kappa, 1]$ and derivative in the same region except for points $1/(2\tilde{\kappa})$ and $1/\tilde{\kappa}$. One can prove that in this case the value  $\norm{ \ket{h(\lambda_1)} - \ket{h(\lambda_2)} } / \abs{\lambda_1 - \lambda_2}$ is upper bounded by $\max_{\lambda} \norm{ d\ket{h(\lambda)}/d\lambda } = \pi/2$. See Section \hyperref[appendix]{Appendix} for detailed proof. }] \label{lemma: continuity}
    The mapping $\lambda \to \ket{h(\lambda)}$ is $O(\kappa)$-Lipschitz. 
    Namely, for any $\lambda_1 \neq \lambda_2$, 
    \begin{align} \label{eq: h(lambda) continuity}
        \norm{ \ket{h(\lambda_1)} - \ket{h(\lambda_2)} } \leq c \kappa \abs{\lambda_1 - \lambda_2},
    \end{align}    
where $c = O(1)$ is a constant.
\end{lemma}

From \eqref{eq: distance between two states} and \eqref{eq: h(lambda) continuity}, we have
\begin{align}
    \norm{ \ket{h(\tilde{\lambda}_k)} - \ket{h(\lambda_j)} } 
    = \sqrt{2(1-\Re\braket{h(\tilde{\lambda}_k)|h(\lambda_j)})}
    \leq c \kappa \abs{\tilde{\lambda}_k - \lambda_j} , 
\end{align}
which yields 
\begin{align} \label{eq: lower bound of inner product between h(lambda_k) and h(lambda_J)}
    \Re\braket{h(\tilde{\lambda}_k)|h(\lambda_j)} \geq 1 - \frac{1}{2} c^2 \kappa^2 (\tilde{\lambda}_k - \lambda_j)^2
    = 1 - \frac{c^2}{2} \frac{\kappa^2}{t_0^2} \delta^2.
\end{align}
With \eqref{eq: lower bound of inner product between h(lambda_k) and h(lambda_J)}, one can prove that 
\begin{align}
    \text{Term 1} &\coloneqq \sum_{j=0}^{N-1} \sum_{k: \abs{\delta} \leq 2\pi}  \abs{\beta_j}^2 \abs{\alpha_{k|j}}^2 \Re{\braket{h(\tilde{\lambda}_k)|h(\lambda_j)} } \\
    & \geq \sum_{j=0}^{N-1} \sum_{k: \abs{\delta} \leq 2\pi}  \abs{\beta_j}^2 \abs{\alpha_{k|j}}^2 - \frac{c^2}{2} \frac{\kappa^2}{t_0^2} \sum_{j=0}^{N-1} \sum_{k: \abs{\delta} \leq 2\pi}  \abs{\beta_j}^2 \abs{\alpha_{k|j}}^2 \delta^2 \\
    & \geq \sum_{j=0}^{N-1} \sum_{k: \abs{\delta} \leq 2\pi}  \abs{\beta_j}^2 \abs{\alpha_{k|j}}^2 - 2 \pi^2 c^2 \frac{\kappa^2}{t_0^2} \label{eq: term 1 upper bound}
\end{align}

To bound Term 2, another fact we need is that $\abs{\alpha_{k|j}}$ is upper bounded when $\delta$ is large, i.e., when the eigenvalue approximation is poor.
In Section \ref{sec: amplitude behavior}, it has been shown numerically that $\abs{\alpha_{k|j}}$ is bounded by 
\begin{align} \label{eq: upper bound for alpha}
    \abs{\alpha_{k|j}} < \frac{8\pi}{\delta^2} \quad \text{if} \quad \abs{\delta} > 2\pi. 
\end{align}
The analytical proof of \eqref{eq: upper bound for alpha} is provided in Appendix \ref{appd: upper bound of alpha_mag}.

Based on \eqref{eq: lower bound of inner product between h(lambda_k) and h(lambda_J)} and \eqref{eq: upper bound for alpha}, we have
\begin{align}
    \text{Term 2} &\coloneqq 
    \sum_{j=0}^{N-1} \sum_{k: \abs{\delta} > 2\pi}  \abs{\beta_j}^2 \abs{\alpha_{k|j}}^2 \Re{\braket{h(\tilde{\lambda}_k)|h(\lambda_j)} } \\
    & \geq 
    \sum_{j=0}^{N-1} \sum_{k: \abs{\delta} > 2\pi}  \abs{\beta_j}^2 \abs{\alpha_{k|j}}^2 - \frac{c^2 \kappa^2}{2t_0^2} \sum_{j=0}^{N-1} \sum_{k: \abs{\delta} > 2\pi}  \abs{\beta_j}^2 \abs{\alpha_{k|j}}^2  \delta^2 \\
    & \geq 
    \sum_{j=0}^{N-1} \sum_{k: \abs{\delta} > 2\pi}  \abs{\beta_j}^2 \abs{\alpha_{k|j}}^2 - \frac{32 \pi^2 c^2 \kappa^2}{t_0^2} \sum_{j=0}^{N-1} \sum_{k: \abs{\delta} > 2\pi}  \abs{\beta_j}^2 \frac{1}{\delta^2} \label{eq: second to last} \\
    & \geq 
    \sum_{j=0}^{N-1} \sum_{k: \abs{\delta} > 2\pi}  \abs{\beta_j}^2 \abs{\alpha_{k|j}}^2 - \frac{4 \pi^2 c^2 \kappa^2}{3t_0^2} \label{eq: term 2 upper bound}
\end{align}
where we used \footnote{See Section \ref{appd: sum over k for poor eigenvalue approximation} for derivation.}
\begin{align} \label{eq: proof in appendix}
    \sum_{j=0}^{N-1} \abs{\beta_j}^2 \sum_{k: \abs{\delta} > 2\pi} \frac{1}{\delta_{k|j}^2} \leq \frac{1}{24}
\end{align}
to get from \eqref{eq: second to last} to \eqref{eq: term 2 upper bound}.

From \eqref{eq: term 1 upper bound}, \eqref{eq: term 2 upper bound}, and \eqref{eq: Re(inner product)}, we have 
\begin{align} \label{eq: Re(inner product) lower bound}
    \Re\{\braket{\Phi_f|\bar{\Phi}_f} \} \geq 1 - \frac{10\pi^2 c^2}{3} \frac{\kappa^2}{t_0^2}.
\end{align}
Plugging \eqref{eq: Re(inner product) lower bound} back into \eqref{eq: distance between two ideal and practical final state, no post selection}, we have completed the proof for claim  \eqref{eq: error bound, no post-selection}.
\begin{align} 
    \norm{ \ket{\Phi_f} -\ket{\bar{\Phi}_f} } \leq \sqrt{\frac{20}{3}} \pi c \frac{\kappa}{t_0} = O(\frac{\kappa}{t_0}). 
\end{align}

\subsection{Proof of \eqref{eq: error bound, post selection to well and ill}: Error Bound in the Solution State with Flag Register Post-Selected in $\{\ket{well}, \ket{ill}\}$} \label{sec: error bound, post-selection onto well and ill}

The post-selection is introduced in Section \ref{sec: algorithm detail}. 
From the ideal and non-ideal final states \eqref{eq: final state, ideal} and \eqref{eq: final state, approximated}, the post-selection onto the $\{\ket{well}, \ket{ill}\}$ subspace of the flag register results in the ideal and non-ideal solution states 
\begin{align}
    \ket{\bar{x}} = \frac{1}{\sqrt{\bar{p}}} \sum_{j=0}^{N-1} \beta_j \ket{u_j} \left[ f(\lambda_j)\ket{well} + g(\lambda_j) \ket{ill} \right],
\end{align}
\begin{align}
    \ket{x} = \frac{1}{\sqrt{p}} P^\dagger \sum_{j=0}^{N-1} \beta_j \ket{u_j} \sum_{k=0}^{T-1} \alpha_{k|j} \ket{k} \left[ f(\tilde{\lambda}_k)\ket{well} + g(\tilde{\lambda}_k) \ket{ill} \right],
\end{align}
where the post-selection success probabilities are
\begin{align} \label{eq: post selection probability, actual}
    \bar{p} = \sum_{j=0}^{N-1} \abs{\beta_j}^2  \left[ f^2(\lambda_j) + g^2(\lambda_j) \right],
\end{align}
and
\begin{align}
    p = \sum_{j=0}^{N-1} \abs{\beta_j}^2 \sum_{k=0}^{T-1} \abs{\alpha_{k|j}}^2 \left[ f^2(\tilde{\lambda}_k) + g^2(\tilde{\lambda}_k) \right].
\end{align}
The inner product between states $\ket{\bar{x}}$ and $\ket{x}$ is thus
\begin{align} \label{eq: fidelity, step 1}
    \braket{x|\bar{x}}
    = \frac{1}{\sqrt{p\bar{p}}} 
     \sum_{j=0}^{N-1} \abs{\beta_j}^2 \sum_{k=0}^{T-1} \abs{\alpha_{k|j}}^2 \left[ f(\tilde{\lambda}_k)f(\lambda_j) + g(\tilde{\lambda}_k) 
 g(\lambda_j) \right].
\end{align}
Note that $\braket{x|\bar{x}}$ is real, hence, 
\begin{align}
    \braket{x|\bar{x}} 
    = 
    \braket{\bar{x}|x}
    =
    \Re\{ \braket{x|\bar{x}} \}
    = 
    \Re\{ \braket{\bar{x}|x} \}.
\end{align}
To simplify the notation, we introduce the following definitions: 
\begin{align}
    f_j \coloneqq f(\lambda_j), \qquad
    \tilde{f}_k \coloneqq f(\tilde{\lambda}_k), \\
    g \coloneqq g(\lambda_j), \qquad
    \tilde{g} \coloneqq g(\tilde{\lambda}_k). 
\end{align}
Furthermore, we treat $\abs{\beta_j}^2$ and $\abs{\alpha_{k|j}}^2$ as distributions and introduce the expectation functions
\begin{align}
    \mathbb{E} (X_j) = \sum_{j=0}^{N-1} \abs{\beta_j}^2  X_j,
\end{align}
\begin{align}
    \mathbb{E} (X_{k,j}) = \sum_{j=0}^{N-1} \abs{\beta_j}^2 \sum_{k=0}^{T-1} \abs{\alpha_{k|j}}^2 X_{k,j},
\end{align}
where $X_j$ and $X_{k,j}$ are random variables.
Therefore, equations \eqref{eq: post selection probability, actual} to \eqref{eq: fidelity, step 1} can be represented as the expectations as 
\begin{align} \label{eq: expectation form of probabilities}
    \bar{p} = \mathbb{E} (f_j^2 + g_j^2), 
    \qquad
    p = \mathbb{E} (\tilde{f}_k^2 + \tilde{g}_k^2), 
    \qquad
     F &= \frac{1}{\sqrt{p\bar{p}}} 
     \mathbb{E} \left( \tilde{f}_k f_j + \tilde{g}_k 
 g_j \right).
\end{align}
With more derivation details included in the Appendix, \eqref{eq: fidelity, step 1} is lower bounded as 
\begin{align}
\braket{x|\bar{x}}  &= \frac{1}{\sqrt{p\bar{p}}} \mathbb{E} \left( \tilde{f}_k f_j + \tilde{g}_k 
 g_j \right)
 = \frac{1 + \mathbb{E} \left( (\tilde{f}_k - f_j) f_j + (\tilde{g}_k-g_j)g_j \right)/\bar{p}}{\sqrt{1+\frac{p-\bar{p}}{\bar{p}}}} \\
 &\geq \left[1 + \mathbb{E} \left[ (\tilde{f}_k - f_j) f_j + (\tilde{g}_k-g_j)g_j \right]/\bar{p}\right] \left( 1-\frac{1}{2}\frac{p-\bar{p}}{\bar{p}} \right) \\
 &\geq 1 - \underbrace{\frac{1}{2\bar{p}} \mathbb{E} \left[(\tilde{f}_k - f_j)^2 + (\tilde{g}_k-g_j)^2 \right] }_\text{Term 1} - \underbrace{\frac{1}{2} \frac{p -\bar{p}}{\bar{p}^2} \mathbb{E} \left[(\tilde{f}_k - f_j)f_j + (\tilde{g}_k-g_j)g_j \right]}_\text{Term 2}. \label{eq: F and Term 1 and Term 2}
\end{align}
Next, the upper bounds of Term 1 and Term 2 are studied to yield the lower bound of $\braket{x|\bar{x}} $.
One important prerequisite for this study is the behavior of the filter functions, which is given in Lemma \ref{lemma: filter function upper bound}, whose proof is provided in the Appendix.
\begin{lemma}[Upper bound of $(\tilde{f}_k - f_j)^2 + (\tilde{g}_k-g_j)^2$] \label{lemma: filter function upper bound}
\begin{align}
    (\tilde{f}_k - f_j)^2 + (\tilde{g}_k-g_j)^2 = O\left(\frac{\kappa^2}{t_0^2} \delta_{k|j}^2 (f_j^2 + g_j^2)\right).
\end{align}
\end{lemma}
Applying Lemma \ref{lemma: filter function upper bound} to the expectation in Term 1 gives
\begin{align}
    \mathbb{E} \left[(\tilde{f}_k - f_j)^2 + (\tilde{g}_k-g_j)^2 \right] 
    = 
    O\left(\frac{\kappa^2}{t_0^2} \mathbb{E} \left[\delta_{k|j}^2 (f_j^2 + g_j^2)\right]\right).
\end{align}
Using \eqref{eq: expectation form of probabilities}, Term 1 becomes
\begin{align} \label{eq: term 1, post selection error analysis}
    \text{Term 1} = 
    \frac{1}{2\bar{p}} \mathbb{E} \left[(\tilde{f}_k - f_j)^2 + (\tilde{g}_k-g_j)^2 \right] 
    = O\left(\frac{\kappa^2}{t_0^2} \frac{ \mathbb{E} \left[\delta_{k|j}^2 (f_j^2 + g_j^2)\right]}{\mathbb{E}(f_j^2 + g_j^2)}\right),
\end{align}
in which the ratio between the expectations is  
\begin{align} \label{eq: ratio between expectations}
    \frac{ \mathbb{E} \left[\delta_{k|j}^2 (f_j^2 + g_j^2)\right]}{\mathbb{E}(f_j^2 + g_j^2)}
    = \frac{ \sum_{j=0}^{N-1} \abs{\beta_j}^2 (f_j^2 + g_j^2) \sum_{k=0}^{T-1} \abs{\alpha_{k|j}^2}\delta_{k|j}^2 }{\sum_{j=0}^{N-1} \abs{\beta_j}^2 (f_j^2 + g_j^2)}.
\end{align}
In the appendix, it is proved that $\mathbb{E}(\delta_{k|j}^2) = O(1)$.
Subsequently,  $\sum_{k=0}^{T-1} \abs{\alpha_{k|j}^2}\delta_{k|j}^2 $ is also $O(1)$. 
Hence, the ratio \eqref{eq: ratio between expectations} is $O(1)$ as it is a weighted sum.
This concludes that Term 1 \eqref{eq: term 1, post selection error analysis} is $O(\kappa^2/t_0^2)$.

Transitioning to the examination of Term 2, first we note that
\begin{align}
    \mathbb{E} &\left[(\tilde{f}_k - f_j)f_j + (\tilde{g}_k-g_j)g_j \right]
    \leq
    \mathbb{E} \left[\abs{ (\tilde{f}_k - f_j)f_j + (\tilde{g}_k-g_j)g_j } \right] \\
    &\leq 
    \mathbb{E} \left[\sqrt{(\tilde{f}_k-f_j)^2 + (\tilde{g}_k-g_j)^2} \sqrt{f_j^2 + g_j^2} \right]
    = O \left( \frac{\kappa}{t_0} \mathbb{E} \left[ \abs{\delta_{k|j}} (f_j^2 + g_j^2) \right] \right)
\end{align}
where we have used the fact that $(\tilde{f}_k - f_j)f_j + (\tilde{g}_k-g_j)g_j$ is not necessary positive, Cauchy-Schwartz inequality, and Lemma \ref{lemma: filter function upper bound} in sequence in the derivation above.
Given that $\mathbb{E}(\abs{\delta_{k|j}}) = O(1)$\footnote{Proof can be found in the Appendix}, we find that
\begin{align} \label{eq: ratio 1 in term 2}
    \frac{\mathbb{E} \left[(\tilde{f}_k - f_j)f_j + (\tilde{g}_k-g_j)g_j \right]}{\bar{p}} 
    = 
    O \left( \frac{\kappa}{t_0} \frac{\mathbb{E} \left[ \abs{\delta_{k|j}} (f_j^2 + g_j^2) \right] }{\mathbb{E}(f_j^2 + g_j^2)} \right)
    = 
    O \left( \frac{\kappa}{t_0} \right)
\end{align}
in Term 2.
The other multiplier of Term 2 is
\begin{align}
    \frac{p-\bar{p}}{\bar{p}}
    &= 
    \frac{\mathbb{E} \left[ (\tilde{f}_k - f_j)^2 + (\tilde{g}_k-g_j)^2 \right] + 2 \mathbb{E} \left[ (\tilde{f}_k - f_j)f_j + (\tilde{g}_k-g_j)g_j \right]}{\mathbb{E}(f_j^2 + g_j^2)} \\
    &= 
    O\left(\frac{\kappa^2}{t_0^2} \frac{\mathbb{E} \left[\delta_{k|j}^2 (f_j^2 + g_j^2)\right]}{\mathbb{E}(f_j^2 + g_j^2)} \right) + O \left( \frac{\kappa}{t_0} \frac{\mathbb{E} \left[ \abs{\delta_{k|j}} (f_j^2 + g_j^2) \right]}{\mathbb{E} \left[(f_j^2 + g_j^2)\right]} \right) \\
    &= 
    O\left(\frac{\kappa^2}{t_0^2} \mathbb{E} \left(\delta_{k|j}^2\right) \right) + O \left( \frac{\kappa}{t_0} \mathbb{E} \left( \abs{\delta_{k|j}} \right)\right)
    = 
    O\left(\frac{\kappa^2}{t_0^2} \right) + O \left( \frac{\kappa}{t_0} \right). \label{eq: last step in (p-pbar)/pbar}
\end{align}
Recall that $t_0 = \Omega(\kappa)$ from Section \ref{sec: amplitude behavior}, leading to the fact that $\kappa/t_0 < 1$, plugging which to \eqref{eq: last step in (p-pbar)/pbar} yields 
\begin{align} \label{eq: ratio 2 in Term 2}
    \frac{p-\bar{p}}{\bar{p}} = O\left( \frac{\kappa}{t_0} \right). 
\end{align}
Term 2 is the product of \eqref{eq: ratio 1 in term 2} and \eqref{eq: ratio 2 in Term 2}, hence, Term 2 is $O(\kappa^2/t_0^2)$. 
At this point, we have proved that both Term 1 and Term 2 are $O(\kappa^2/t_0^2)$. 
Subsequently from \eqref{eq: F and Term 1 and Term 2},  
\begin{align}
\braket{x|\bar{x}} = 1- O\left(\frac{\kappa^2}{t_0^2}\right) .
\end{align}
Using \eqref{eq: distance between two states}, we have
\begin{align} 
    \norm{\ket{x}  - \ket{\bar{x}}}
    = \sqrt{2(1-\Re\{\braket{x|\bar{x}}\})} 
    = O\left(\frac{\kappa}{t_0}\right).
\end{align}

\subsection{Proof of \eqref{eq: error bound, post selection to well}: Error Bound in the Solution State with Flag Register Post-Selected in $\{\ket{well}\}$}
To utilize the analysis performed in Sections \ref{sec: error bound, no post-selection} and \ref{sec: error bound, post-selection onto well and ill}, we regard the process of obtaining the solution state as first post selecting the flag register on the subspace of $\{\ket{well}, \ket{ill}\}$, followed by another post selection on to $\{\ket{well}\}$. 
We denote these two post selection operators as $PS_1$ and $PS_2$.
As the phase register does not entangle with other registers in the final state, in this section, we omit it in all representations.

When $\ket{b}$ stays in the well-conditioned subspace of $A$, i.e., $\ket{b}$ is a linear combination of eigenvalues of $A$ whose eigenvalues are in $\left[\frac{1}{\tilde{\kappa}} , 1 \right]$, the ideal final state $\ket{\bar{\Phi}_f}$ would have no $\ket{ill}$ and $\ket{nothing}$ components in the flag register
\begin{align}
    \ket{\bar{\Phi}_f} = A^{-1} \ket{b} \ket{well}.
\end{align}
Thus, both post-selections have the success probability of 1.
\begin{align} \label{eq: bar x1}
    \ket{\bar{x}_1} \coloneqq PS_1  \ket{\bar{\Phi}_f} = A^{-1} \ket{b}\ket{well},
\end{align}
\begin{align}
    \ket{\bar{x}_2} \coloneqq PS_2 \ket{\bar{x}_1} = A^{-1} \ket{b}\ket{well},
\end{align}
and the solution state is 
\begin{align} \label{eq: bar x}
    \ket{\bar{x}} = A^{-1} \ket{b}.
\end{align}
The non-ideal case final state contains components in all three subspaces of the flag register spanned by its basis states $\{\ket{well}\}$, $\{\ket{ill}\}$, $\{\ket{nothing}\}$\footnote{Rigorously speaking, it is possible that some subspace has a all-zero vector in register $I$, in which case using ket notation is not precise. We note that this case does not break the proof and disregard it in the analysis}: 
\begin{align}
    \ket{\Phi_f} = \mathbf{x}_f^w \ket{well} + \mathbf{x}_f^i \ket{ill} + \mathbf{x}_f^n \ket{nothing}.
\end{align}
$PS_1$ succeeds with the probability $p_1 = \norm{\mathbf{x}_f^w}^2 + \norm{\mathbf{x}_f^i }^2 $ and yields 
\begin{align} \label{eq: x1}
    \ket{x_1} \coloneqq PS_1{\ket{\Phi_f}} 
    = \mathbf{x}_1^w \ket{well} + \mathbf{x}_1^i \ket{ill}, 
    \quad
    \mathbf{x}_1^w = \mathbf{x}_f^w / \sqrt{p_1}, 
    \quad
    \mathbf{x}_1^i = \mathbf{x}_f^i / \sqrt{p_1}
\end{align}
Similarly, $PS_2$ succeeds with the probability $p_2 = \norm{\mathbf{x}_1^w}^2$ and yields 
\begin{align} \label{eq: x2}
    \ket{x_2} \coloneqq PS_2{\ket{x_1}} 
    = \ket{x} \ket{well}, 
    \quad
    \ket{x} = \mathbf{x}_1^w / \sqrt{p_2}.
\end{align}

From Section \ref{sec: error bound, post-selection onto well and ill}, we know that
\begin{align} \label{eq: x1 inner product}
    \Re \{\braket{x_1|\bar{x}_1} \} = 1- O \left(\frac{\kappa^2}{t_0^2}\right) 
\end{align}
holds in general. 
Within the context of this section, i.e., 
\eqref{eq: bar x1}, \eqref{eq: bar x}, \eqref{eq: x1}, 
\begin{align} \label{eq: x1 inner product with b in A's well conditioned subspace}
    \braket{x_1|\bar{x}_1} = (\mathbf{x}_1^w)^\dagger \ket{\bar{x}}.
\end{align}
Combining \eqref{eq: x1 inner product} and \eqref{eq: x1 inner product with b in A's well conditioned subspace} gives
\begin{align} \label{eq: x1 x1 bar inner product}
    \Re \{\braket{x_1|\bar{x}_1} \} 
    = 
    \Re \{ (\mathbf{x}_1^w)^\dagger \ket{\bar{x}} \}
    =
    1- O \left(\frac{\kappa^2}{t_0^2}\right).
\end{align}
Similar to previous subsections, we rely on \eqref{eq: distance between two states} and prove the upper bound for $\norm{\ket{x} - \ket{\bar{x}}}$ by finding the lower bound for $\Re\{\braket{x|\bar{x}}\}$.
The latter is given as
\begin{align} \label{eq: x x bar inner product}
    \Re\{\braket{x|\bar{x}}\}
    = 
    \frac{1}{\sqrt{p_2}} \Re \{(\mathbf{x}_1^w)^\dagger \ket{\bar{x}}\}
    = \frac{1}{\sqrt{p_2}} - O \left(\frac{\kappa^2}{t_0^2}\right)
    > 1 - O \left(\frac{\kappa^2}{t_0^2}\right),
\end{align}
where we have used \eqref{eq: x2}, \eqref{eq: x1 x1 bar inner product}, and the fact that $p_2 < 1$.
Physically, \eqref{eq: x x bar inner product} means that the second post-selection amplifies the error by $1/\sqrt{p_2}$, however, this does not affect the overall error scaling.
Plugging \eqref{eq: x x bar inner product} into \eqref{eq: distance between two states} completes the proof of \eqref{eq: error bound, post selection to well}. 

\subsection{Summary}
To summarize the error analysis, all claims are proved by finding the lower bound for the real part of the inner product between the ideal and non-ideal states. 
The inner product for the final state is studied by separating the contributions from good and bad eigenvalue approximations.
The post-selection error to $\{\ket{well}, \ket{ill}\}$ subspace is bounded by expressing the division by the success probability as a weighted sum.
When projecting to the $\{\ket{well}\}$ subspace, the proof is finalized by straightforwardly applying the previous analysis.  
In practice, the error bound provides the instruction on the selection on $t_0$ to achieve a prescribed target error level $\epsilon$: choosing $t_0 = O(\kappa/\epsilon)$. 
Note that this choice adheres to the choice we discussed in Section \ref{sec: amplitude behavior}: $t_0 = \gamma (2\pi T)$, $T \geq \kappa/\gamma + 1$ with $\gamma < 1/2$.

\section{Conclusion}
A detailed error analysis of the HHL algorithm is presented in this work.
Besides offering corrections to mistakes, compared with the supplementary material of \cite{Harrow_2009}, this analysis covers derivation details with explicit expressions to be more approachable.
The connection between the Hamiltonian simulation duration $t_0$ and the clock register size $T$ is also established as a result of the analysis of the amplitudes after QPE, without assuming an infinite $T$ as in \cite{Harrow_2009}.
We hope this work would be useful to those who try to discover improved version of the HHL algorithm when an error analysis is needed for the derived algorithm.

\clearpage
\appendix

\section{The inner product between ideal and non-ideal final states} \label{appd: inner product}
Based on the definition of QPE $P$ in Section \ref{sec: algorithm detail}, it operating on $\ket{\Phi_f}$ \eqref{eq: final state, ideal} would yield 
\begin{align}
    P \ket{\bar{\Phi}_f} = \sum_{j=0}^{N-1} \beta_j \ket{u_j} \sum_{k=0}^{T-1} \alpha_{k|j} \ket{k} \ket{h(\lambda_j)}. 
\end{align}
Hence,
\begin{align}
     \braket{\Phi_f|\bar{\Phi}_f} 
    &= \braket{\Phi_f | P^\dagger  P |\bar{\Phi}} \\
    &= \left( \sum_{j=0}^N \beta_j^* \bra{u_j}
            \sum_{k=0}^{T-1} \alpha_{k|j}^* \bra{k} \bra{h(\tilde{\lambda}_k)} \right)
    \left( \sum_{j^\prime=0}^{N-1} \beta_{j^\prime} \ket{u_{j^\prime}} \sum_{k^\prime=0}^{T-1} \alpha_{k^\prime|j^\prime} \ket{k^\prime} \ket{h(\lambda_{j^\prime})} \right) \\
    & = \sum_{j=0}^{N-1} \sum_{k=0}^{T-1} \abs{\beta_j}^2 \abs{\alpha_{k|j}}^2 \braket{h(\tilde{\lambda}_k)|h(\lambda_j)}
\end{align}

\section{The mapping $\lambda \to \ket{h(\lambda)}$ is $O(\kappa)$-Lipschitz.} \label{appd: continuity of h(lambda)}

\subsection{Model $\ket{h(\lambda)}$ as a Vector-Valued Function}
Suppose a vector-valued function 
\begin{align}
    \mathbf{f}(t) =  
    \begin{bmatrix}
        x(t)\\
        y(t)\\
        z(t)
    \end{bmatrix},
\end{align}
is defined in $t \in [t_L, t_R]$, and each component $x(t), y(t)$, and $z(t)$ is differentiable in $(t_L, t_R)$ except at two points $t_1$ and $t_2$ with $t_L < t_1 < t_2 < t_R$.
We aim to find a upper bound of the value 
\begin{align} \label{eq: slope squared vector-valued function}
    \frac{\norm{\mathbf{f}(t_l) - \mathbf{f}(t_r)}^2}{\abs{t_l - t_r}^2 } = 
    \frac{(x(t_l)-x(t_r))^2}{(t_l - t_r)^2} +
    \frac{(y(t_l)-y(t_r))^2}{(t_l - t_r)^2} + 
    \frac{(z(t_l)-z(t_r))^2}{(t_l - t_r)^2}
\end{align}
for arbitrary two points $t_l < t_r$ from $[t_L, t_R]$.

If both $t_l$ and $t_r$ are contained in a single domain from the set $S= \{ [t_L, t_1), (t_1, t_2), (t_2, t_R] \}$, we apply the mean-value theorem and obtain that
\begin{align}
    \frac{\norm{\mathbf{f}(t_l) - \mathbf{f}(t_r)}^2}{\abs{t_l - t_r}^2 } &= 
    \frac{(x(t_l)-x(t_r))^2}{(t_l - t_r)^2} +
    \frac{(y(t_l)-y(t_r))^2}{(t_l - t_r)^2} + 
    \frac{(z(t_l)-z(t_r))^2}{(t_l - t_r)^2} \\
    &= [x^\prime(c_1)]^2 + [y^\prime(c_2)]^2 + [z^\prime(c_3)]^2, \label{eq: vector-valued function mean value theorem}
\end{align}
where $c_1, c_2$, and $c_3$ are some constants in $(t_l, t_r)$, and the superscript prime denotes derivative.
Equation \eqref{eq: vector-valued function mean value theorem} suggests that 
\begin{align} \label{eq: slope squared vector-valued function upper bound}
    \frac{\norm{\mathbf{f}(t_l) - \mathbf{f}(t_r)}^2}{\abs{t_l - t_r}^2 } \leq 
    \max_{t \in (t_l, t_r)} [x^\prime(t)]^2 + 
    \max_{t \in (t_l, t_r)} [y^\prime(t)]^2 +
    \max_{t \in (t_l, t_r)} [z^\prime(t)]^2.    
\end{align}

Now let us consider the case when $t_l$ and $t_r$ belong to different domains from the set $S$. 
Suppose $t_l$ and $t_r$ are from two adjacent domains in $S$. 
We symbolize the disconnecting point of the two domains as $t_c$. 
Then applying the mean value theorem in regions $(t_l, t_c)$ and $(t_c, t_r)$ to $x(t)$ gives 
\begin{align}
    x(t_r) - x(t_c) = (t_r-t_c) x^\prime(c_1) ,
\end{align}
\begin{align}
    x(t_c) - x(t_l) = (t_c-t_l) x^\prime(c_2) .
\end{align}
As a result, 
\begin{align}
    \abs{ x(t_r) - x(t_l) } &=
    \abs{ x(t_r) - x(t_c) + x(t_c) - x(t_l) }\\
    &\leq \abs{ x(t_r) - x(t_c)} + \abs{x(t_c) - x(t_l) } \\
    &= \abs{(t_r-t_c) x^\prime(c_1)} + \abs{(t_c-t_l) x^\prime(c_2)} \\
    &= (t_r-t_c) \abs{x^\prime(c_1)} + (t_c-t_l) \abs{x^\prime(c_2)} \\ 
    &\leq (t_r-t_c) \max_{t \in (t_c, t_r)}{\abs{x^\prime(t)}} + (t_c-t_l) \max_{t \in (t_l, t_c)}{\abs{x^\prime(t)}}\\
    &\leq (t_r-t_l) \max_{t \in (t_l, t_r)}{\abs{x^\prime(t)}}. 
\end{align}
In other words, 
\begin{align} \label{eq: disconnected domains, x}
    \frac{\abs{ x(t_r) - x(t_l)}^2}{(t_r-t_l)^2} \leq \max_{t \in (t_l, t_r)} [x^\prime(t)]^2.
\end{align}
Same procedure applies to $y(t)$ and $z(t)$. 
Hence,
\begin{align} \label{eq: disconnected domains, y}
    \frac{\abs{ y(t_r) - y(t_l)}^2}{(t_r-t_l)^2} \leq \max_{t \in (t_l, t_r)} [y^\prime(t)]^2.
\end{align}
\begin{align} \label{eq: disconnected domains, z}
    \frac{\abs{ z(t_r) - z(t_l)}^2}{(t_r-t_l)^2} \leq \max_{t \in (t_l, t_r)} [z^\prime(t)]^2.
\end{align}
Equation \eqref{eq: slope squared vector-valued function} together with \eqref{eq: disconnected domains, x}, \eqref{eq: disconnected domains, y}, \eqref{eq: disconnected domains, z} indicates that the upper bound equation \eqref{eq: slope squared vector-valued function upper bound} still holds in this case.

Following a similar procedure, one can prove that if $t_l$ and $t_r$ are not from adjacent domains in $S$, equation \eqref{eq: slope squared vector-valued function upper bound} also holds.

In summary, for arbitrary two points $t_l < t_r$ in domain $[t_L, t_R]$, one upper bound for the value of interest expressed in equation \eqref{eq: slope squared vector-valued function} is given as equation \eqref{eq: slope squared vector-valued function upper bound}. 

\subsection{The Study for the Mapping $\lambda \to \ket{h(\lambda)}$}
Now let us study the mapping $\lambda \to \ket{h(\lambda)}$.
It is clear from the definition \eqref{eq: h(lambda) definition} that $\ket{h(\lambda)}$ can be modeled as a vector-valued function defined on domain $[0,1]$, where the components $\sqrt{1-f^2(\lambda)-g^2(\lambda)}$, $f(\lambda)$, and $g(\lambda)$ are differentiable on $(0, \frac{1}{2\tilde{\kappa}}) \cup (\frac{1}{2\tilde{\kappa}},\frac{1}{\tilde{\kappa}}) \cup (\frac{1}{\tilde{\kappa}}, 1)$. 
Thus, the study from the last subsection applies to $\ket{h(\lambda)}$, leading to
\begin{align}
    \frac{\norm{\ket{h(\lambda_2)} - \ket{h(\lambda_1)}}^2}{\abs{\lambda_2 - \lambda_1}^2 } &\leq 
    \max_{\lambda \in (0, 1)} \left[\frac{d}{d\lambda} \sqrt{1-f^2(\lambda)-g^2(\lambda)} \right]^2 \nonumber \\
    &\quad + 
    \max_{\lambda \in (0, 1)} \left[\frac{df(\lambda)}{d\lambda } \right]^2 +
    \max_{\lambda \in (0, 1)} \left[\frac{dg(\lambda)}{d\lambda } \right]^2 \\
    & = \left[ \frac{\tilde{\kappa}}{2\sqrt{3}} \right]^2 + \left[ \frac{\pi}{2} \tilde{\kappa} \right]^2 + \left[ \frac{\pi}{2} \tilde{\kappa} \right]^2 
    = \frac{6\pi^2 + 1}{12} \tilde{\kappa}^2
\end{align}
for any $\lambda_1 \neq \lambda_2$ in $[0,1]$. 
Hence, 
\begin{align} \label{eq: one step before}
    \norm{\ket{h(\lambda_2)} - \ket{h(\lambda_1)}} \leq  \sqrt{\frac{6\pi^2 + 1}{12}} \tilde{\kappa} \abs{\lambda_2 - \lambda_1}.
\end{align}
Equation \eqref{eq: one step before} together with \eqref{eq: kappa_t and kappa} gives that  
\begin{align} 
    \norm{\ket{h(\lambda_2)} - \ket{h(\lambda_1)}} = O(\kappa)  \abs{\lambda_2 - \lambda_1},
\end{align}
which completes the proof of Lemma \ref{lemma: continuity}. \qed

Additional note: We are not attempting to find the best upper bound for $\norm{\ket{h(\lambda_2)} - \ket{h(\lambda_1)}}/\abs{\lambda_2 - \lambda_1}$ in the proof, rather simply prove that it is bounded by $O(\kappa)$. 
Tighter upper bound than the one in \eqref{eq: one step before} may be found, but we note that the factor $\pi/2$ as claimed in \cite{Harrow_2009} is not correct for certain values of $\kappa$.

\section{Simplified expressions of $\alpha_{k|j}$ and $\abs{\alpha_{k|j}}$} \label{appd: amplitude}
\begin{align} \label{eq: alpha}
    \alpha_{k|j} 
    &= \frac{\sqrt{2}}{T}\sum_{\tau=0}^{T-1} \sin \left[\frac{\pi}{T} \left(\tau + \frac{1}{2} \right) \right] e^{i \left(\lambda_j - \frac{2\pi}{t_0} k \right) \frac{t_0}{T} \tau} \\
    &= \frac{\sqrt{2}}{T}\sum_{\tau=0}^{T-1} \sin \left[\frac{\pi}{T} \left(\tau + \frac{1}{2} \right) \right] e^{i \tau \delta/T} \\
    &= \frac{\sqrt{2}}{T} \frac{1}{2i} \sum_{\tau=0}^{T-1} \left[ e^{i\frac{\pi}{T} \left(\tau + \frac{1}{2} \right)} - e^{-i\frac{\pi}{T} \left(\tau + \frac{1}{2} \right)}   \right] e^{i \delta \tau/T} \\
    &= \frac{\sqrt{2}}{T} \frac{1}{2i} \left[ e^{i \frac{\pi}{2T}} \sum_{\tau=0}^{T-1}  e^{i \tau \left( \delta + \pi \right)/T} - e^{-i \frac{\pi}{2T}} \sum_{\tau=0}^{T-1} e^{i \tau \left( \delta-\pi \right)/T}  \right] \\
    &= \frac{\sqrt{2}}{T} \frac{1}{2i} \left[ e^{i \frac{\pi}{2T}}
    \frac{1+e^{i \delta}}{1-e^{i\left( \delta+\pi\right)/T}} - e^{-i\frac{\pi}{2T}} \frac{1+e^{i \delta}}{1-e^{i \left(\delta-\pi \right)/T}}  \right] \\
    &= \frac{\sqrt{2}}{T} \frac{1}{2i} \left(1+e^{i\delta}\right) e^{-i \frac{\delta}{2T}} \left[ 
    \frac{1}{e^{-i \frac{\delta+\pi}{2T}} - e^{i \frac{\delta+\pi}{2T}}} - \frac{1}{e^{-i\frac{\delta-\pi}{2T} }-e^{i \frac{\delta-\pi}{2T}}} \right] \\
    &= \frac{\sqrt{2}}{4T} \left(1+e^{i\delta}\right) e^{-i \frac{\delta}{2T}} \left[ 
    \frac{1}{\sin{\left(\frac{\delta+\pi}{2T}\right)}} - \frac{1}{\sin{\left(\frac{\delta-\pi}{2T}\right)}} \right] \\
    &= \frac{\sqrt{2}}{2T} e^{i \frac{\delta}{2} \left(1-\frac{1}{T}\right)} \cos{\left(\frac{\delta}{2}\right)}  
    \frac{\sin{\left(\frac{\delta-\pi}{2T}\right)} - \sin{\left(\frac{\delta+\pi}{2T}\right)}}{\sin{\left(\frac{\delta+\pi}{2T}\right)}\sin{\left(\frac{\delta-\pi}{2T}\right)}} \\
    &= -\frac{\sqrt{2}}{T} e^{i\frac{\delta}{2}\left(1-\frac{1}{T}\right)}  \sin{\left(\frac{\pi}{2T}\right)}
    \frac{\cos{(\frac{\delta}{2T})} \cos(\frac{\delta}{2})}{\sin{\left(\frac{\delta+\pi}{2T} \right)}\sin{\left(\frac{\delta - \pi}{2T} \right)}} .
\end{align}
As a result, 
\begin{align} \label{eq: amp mag}
    \abs{\alpha_{k|j}} = \frac{\sqrt{2}}{T}   \sin{\left(\frac{\pi}{2T}\right)}
    \frac{ \abs*{\cos{(\frac{\delta}{2T})} \cos(\frac{\delta}{2})}}{\abs*{\sin{\left(\frac{\delta+\pi}{2T} \right)}\sin{\left(\frac{\delta - \pi}{2T} \right)}}} .
\end{align}

\section{The upper bound for $\abs{\alpha_{k|j}}$ when $\abs{\delta} > 2\pi$} \label{appd: upper bound of alpha_mag}
From Section \ref{sec: amplitude behavior} we have learnt that $\abs{\delta}$ should be upper bounded by $\pi T$.
As $\abs{\alpha}$ is an even function of $\delta$, it suffices to only analyze the case when $\delta > 0$. 
Hence, the proper bound for $\delta$ to study the poor eigenvalue approximation is $2\pi < \delta < \pi T$. 
We use this bound for $\delta$ thereafter in this section.

Using $\sin x < x$ to the sine function in the numerator, and $\sin x > x- \frac{1}{6} x^3$ to the sine functions in the denominator, and $\abs{\cos x} < 1$ to $\cos (\frac{\delta}{2})$, after some simplifications, we have 
\begin{align}
    \abs{\alpha_{k|j}} \leq 2 \sqrt{2} \pi \cos \left(\frac{\delta}{2T}\right) \frac{1}{\delta^2 - \pi^2}  \frac{1}{1-\frac{\delta^2 + \pi^2}{12T^2}}
\end{align}
Using $\cos x \leq 1 - \frac{1}{2}x^2 + \frac{1}{24} x^4$ and some manipulations, the above is further upper bounded by 
\begin{align}
    \abs{\alpha_{k|j}} \leq \frac{2 \sqrt{2} \pi}{\delta^2} \frac{1 - \frac{\delta^2}{8T^2} + \frac{\delta^4}{384T^4}}{1 - \frac{\pi^2}{\delta^2} - \frac{\delta^2}{12T^2} + \frac{\pi^4}{12 \delta^2 T^2}} 
\end{align}
Next, we prove that 
\begin{align} \label{eq: delta and T bounded by constant}
    \frac{1 - \frac{\delta^2}{8T^2} + \frac{\delta^4}{384T^4}}{1 - \frac{\pi^2}{\delta^2} - \frac{\delta^2}{12T^2} + \frac{\pi^4}{12 \delta^2 T^2}}  \leq \sqrt{2} .
\end{align}
Let us define 
\begin{align}
    a \coloneqq \frac{\delta}{\pi T},
\end{align}
then $a \in (\frac{2}{T}, 1]$.
Proving \eqref{eq: delta and T bounded by constant} amounts to proving 
\begin{align} \label{eq: 6th order poly}
    \frac{\sqrt{2}}{T} \leq 
    -\frac{\pi^4}{384} a^6 + \left(\frac{1}{8} - \frac{\sqrt{2}}{12}\right) \pi^2 a^4 + (\sqrt{2}-1) a^2
\end{align}

\section{Proof of \eqref{eq: proof in appendix}} \label{appd: sum over k for poor eigenvalue approximation}

\begin{align}
    \sum_{k: \abs{\delta_{k|j}} \geq 2\pi} \frac{1}{\delta_{k|j}^2} = 
    \frac{1}{4\pi^2} \sum_{k: \abs{k - \frac{\lambda_j t_0}{2\pi}} \geq 1}  \frac{1}{ \abs{k - \frac{\lambda_j t_0}{2\pi}} ^2 }
\end{align}
As $\frac{\lambda_j t_0}{2\pi}$ may not be an integer, 
\begin{align}
    \abs*{k - \frac{\lambda_j t_0}{2\pi}} \geq 1 
\end{align}
requires that
\begin{align}
    k = \ceil*{ \frac{\lambda_j t_0}{2\pi} } + m, \quad m = 1, 2, \cdots, T - 1 - \ceil*{ \frac{\lambda_j t_0}{2\pi} }.
\end{align}
Thus, 
\begin{align}
    \frac{1}{ \abs*{k - \frac{\lambda_j t_0}{2\pi}} } \leq \frac{1}{m}, 
\end{align}
which results in 
\begin{align}
    \sum_{k: \abs{\delta_{k|j}} \geq 2\pi} \frac{1}{\delta_{k|j}^2} \leq
    \frac{1}{4\pi^2} \sum_{m=1}^{T - 1 - \ceil*{ \frac{\lambda_j t_0}{2\pi}}}  \frac{1}{ m^2 } 
    <
    \frac{1}{4\pi^2} \sum_{m=1}^\infty  \frac{1}{ m^2 } = \frac{1}{24}, 
\end{align}
where we used the fact that
\begin{align}
    \sum_{m=1}^\infty  \frac{1}{ m^2 } = \frac{\pi^2}{6}.
\end{align}

In summary, 
\begin{align}
    \sum_{j=0}^{N-1} \sum_{k: \abs{\delta_{k|j}} \geq 2\pi}  \abs{\beta_j}^2 \frac{1}{\delta_{k|j}^2} 
    < \frac{1}{24} \sum_{j=0}^{N-1} \abs{\beta_j}^2 = \frac{1}{24}, 
\end{align}
as the amplitudes $\beta_j$ satisfy $\sum_{j=0}^{N-1} \abs{\beta_j}^2 = 1$.

\bibliographystyle{unsrtnat}
\bibliography{references}  






\end{document}